\documentclass{article}

\usepackage{arxiv}

\usepackage[utf8]{inputenc} 
\usepackage[T1]{fontenc}    
\usepackage{hyperref}       
\usepackage{url}            
\usepackage{booktabs}       
\usepackage{amsmath} 
\usepackage{cite} 

\usepackage{graphicx}
\usepackage{nicefrac}       
\usepackage{microtype}      
\usepackage{lipsum}
\usepackage{graphicx}
\graphicspath{ {./images/} }

\title{Community detection on directed networks with missing edges}

\author{
 Nicola Pedreschi \\
  Mathematical Institute\\
  University of Oxford\\
  Oxford, UK \\
  \texttt{nicola.pedreschi@maths.ox.ac.uk} \\
   \And
 Renaud Lambiotte \\
  Mathematical Institute\\
  University of Oxford\\
  Oxford, UK \\
  \And
 Alexandre Bovet \\
  Department of Mathematical Modeling and Machine Learning\\
  University of Zurich\\
  Zurich, CH \\
}

\begin{document}

\maketitle
\begin{abstract}
Identifying significant community structures in networks with incomplete data is a challenging task, as the reliability of solutions diminishes with increasing levels of missing information. However, in many empirical contexts, some information about the uncertainty in the network measurements can be estimated. In this work, we extend the recently developed Flow Stability framework, originally designed for detecting communities in time-varying networks, to address the problem of community detection in weighted, directed networks with missing links. Our approach leverages known uncertainty levels in nodes' out-degrees to enhance the robustness of community detection. Through comparisons on synthetic networks and a real-world network of messaging channels on the Telegram platform, we demonstrate that our method delivers more reliable community structures, even when a significant portion of data is missing.
\end{abstract}

%
\vspace{2pc}

\section{Introduction}

Networks are widely used to represent connections in complex systems, such as technological, biological, and social systems \cite{Barrat_weighted_complex,strogatz2001exploring}. Over the past two decades, empirical studies have, in fact, extensively measured and described structures like the Internet \cite{web}, road and airline networks \cite{colizza}, friendship networks \cite{barrat2014measuring,face-to-face}, and biological networks \cite{Towlson6380,bassett2017networkneuro,bullmore2009complex}. 
Analogously to any other empirical discipline, an issue that can occur in applied network science studies is data unreliability due to experimental measurement errors \cite{yeast,krista,err_social}.  Strategies to account for network measurement errors have been put forward over the years, including approaches aimed at predicting missing nodes or edges and at name disambiguation \cite{newman_rich_noisy,Newman_bayesian,newman_weighted}. Combining these approaches can create hybrid algorithms for resampling and estimating network structures.

Among the other network analyses that can be affected by network measurement errors, community detection methods \cite{fortunato2016community,leicht2008community,Blondel_2008} that ignore data uncertainties can overfit, leading to inaccurate classifications of nodes' community assignment \cite{ghasemian_overfit,ghasemian_linkPred,peixoto_reconstruct} (See schema in Figure \ref{fig:schema}.a). To address this, we can compare detected communities with null models or test their robustness under random perturbations \cite{stat_sig_comms,fortuLanciRadi}. As an example, in flow-based community detection methods like the map equation, Bayesian approaches can help in reconstructing the ground truth community structure of a synthetic network, where a randomly selected fraction of the original edges have been removed from the network, detecting and re-inserting into the system the missing links \cite{jelena_unweighted,Jelena_weighted}. 

In this paper, we extend another flow-based community detection method, the recently developed framework of Flow Stability \cite{doi:10.1126/sciadv.abj3063}, to incorporate information on experimental errors in weighted, directed networks in order to reconstruct an accurate community structure of the original network. Flow Stability was originally designed to detect communities in temporal networks \cite{holme2015modern,Holme:2012,NaokiRenaud,ghasemian2016detectability,rossetti2018community}. This method considers random walks restricted by edge activation times \cite{Lambiotte2023}, clustering nodes based on diffusion patterns over time. It generalizes the framework of Markov Stability \cite{renaud_mauricio,Markov_st} from static to temporal networks, uncovering dynamic communities and different scales without temporal aggregation.
In \cite{Bovet_telegram}, Flow Stability (FS) is adapted to detect communities at different, relevant spatial scales in static, directed networks. The method was then used to uncover the organization and influence relations of different communities in a network of far-right Telegram channels and group chats. 

Here, we build on this approach by incorporating the estimates of experimental errors on the ``measurement'' of the network under study into the equations of FS. 
Our novel method is named $\Delta$ Flow Stability ($\Delta$FS). 
In particular, we consider the case in which an estimate of the experimental error on the out-strength of nodes is known. Such choice is motivated by the fact that in social media systems, users can often delete their own messages or messages in channels they control (such as in Telegram or WhatsApp).
When modeling such systems as a network of channels, edges usually represent links posted in messages that allow users to go from one channel to another. When researchers collect messages from these channels, messages deleted before the collection date are no longer accessible. However, it is often possible to estimate the number of deleted messages, for example, by comparing the total number of messages sent in the channel with the actual number of collected messages. 
This estimate can then be used to estimate the number of missing out going edges in the network.

We test our method on synthetically generated networks, generated with a Stochastic Block Model (SBM) \cite{HOLLAND1983109,Tiago_PhysRevE.85.056122,AurelienDec} first assessing the capability of $\Delta$FS to recover the original partition of the network, in comparison to regular FS, when an increasing fraction of randomly selected edges is removed from the original network. We then perform a similar experiment on the Telegram dataset. Finally, we use available information in the Telegram dataset, that allows us to estimate the experimental error on the out-strength of individual nodes, and use it to reconstruct a network that varies substantially in comparison to what is found with FS. By incorporating these uncertainties, our approach reconstructs the community structure, revealing the organization and influence relations of upstream, core, and downstream communities. We identify important clusters of channels and external links that couldn't be observed otherwise, and discuss directions for future work.

\section{$\Delta$ Flow Stability}
\begin{figure}[h!]
    \centering
    \includegraphics[width=0.5\textwidth]{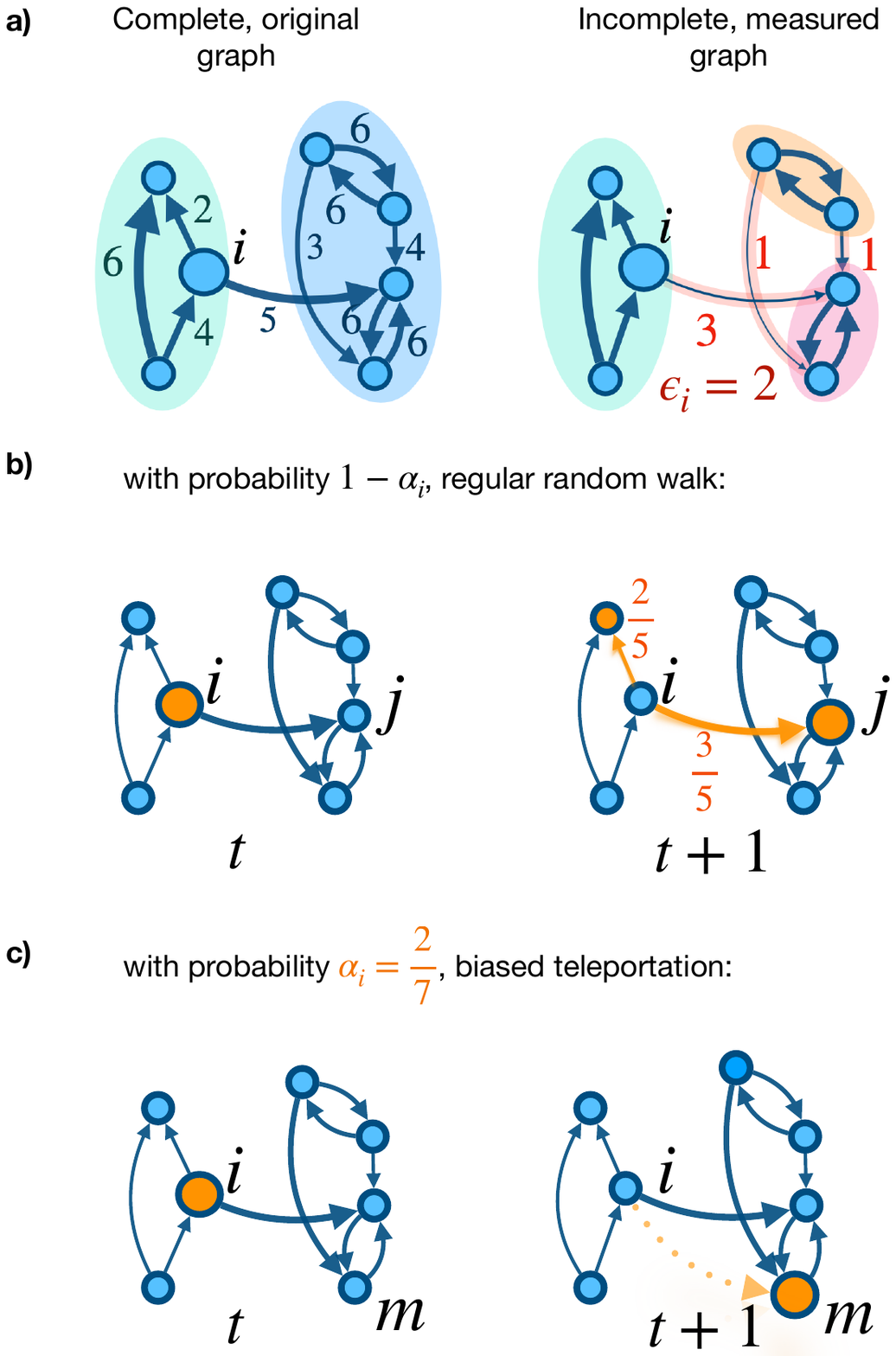}
    \caption{\textbf{$\Delta$ Flow Stability:} \textbf{a)} Illustration of the original, or complete, weighted and directed graph (on the left) characterized by two communities, and the experimental measurement (the incomplete graph, the object of our analysis, on the right) of the graph where some of the edges have not been observed and thus the weights of some connections are changed (in red) w.r.t. the original graph. The missing edges affect the community detection, as the graph now appears to have three communities instead of the original two; note how $2$ edges leaving node $i$ are missing in the measured graph, leading to an overall error $\epsilon_i=2$ on node $i$'s \emph{measured} out-strength $s^i_\text{out}$. \textbf{b)} With probability $1-\alpha_i$, a random walker on node $i$, at any given time step $t$, follows the outgoing edges of $i$ (regular random walk). \textbf{c)} With probability $\alpha_i=\frac{\epsilon_i}{\epsilon_i+s^i_\text{out}}$, instead, the random walker is \emph{teleported} to any other node in the network.}
    \label{fig:schema}
\end{figure}

Flow Stability (FS) was first introduced in \cite{doi:10.1126/sciadv.abj3063} with the aim of detecting communities in time varying (temporal) networks. The core idea of FS is to cluster the covariance matrix of two random walk processes: a forward process, where the diffusion unfolds along the chronological ordering of the edges of the temporal network; and a backward process, where the random walkers follow the edges in reverse chronological order. In both processes, the clustering of the covariance matrix assigns two nodes $i$ and $j$ to the same cluster if the probability that two random walkers, each initially positioned on one of the two nodes, are found on the same node at time $t$ is high. 
In \cite{Bovet_telegram}, authors introduce the adaptation of FS to static, directed networks. In this setup, FS thus allows to cluster together nodes that share a similar \emph{function} in the system under study in terms of diffusive processes unfolding onto the network. Nodes with more outward than incoming connections tend to be clustered together, as they represent ``sources" of the diffusive process and are therefore referred to as \emph{upstream} nodes or clusters of nodes. On the other hand, nodes with an abundance of incoming connections and few outward links correspond to ``sinks" of the diffusive process, thus named \emph{downstream} clusters of nodes. Nodes with an equal number of outward and inward connections are neither sources nor sinks of the diffusion, instead they typically correspond to nodes with high values of in- and out-strength and are therefore referred to as \emph{core} nodes.

The key ingredients in the adaptation of Flow Stability to analyze static, weighted and directed networks, are, similarly to FS for temporal networks, two transition matrices for two continuous time random walk processes: one for the forward process $\mathbf{T}(t)_\text{f}=e^{-t\mathbf{L}_\text{f}}$, following the direction of the edges, one for the backward process $\mathbf{T}(t)_\text{b}=e^{-t\mathbf{L}_\text{b}}$, i.e., following the reverse direction of the edges. Forward and backward Laplacians are given by $\mathbf{L}_\text{f}=\mathbf{I}-\mathbf{D}_\text{out}^{-1}\mathbf{A_\text{f}}$ and $\mathbf{L}_\text{b}=\mathbf{I}-\mathbf{D}_\text{in}^{-1}\mathbf{A_\text{b}}^\textsf{T}$. $\mathbf{A_\text{f}}$ and $\mathbf{A_\text{b}}$ are the forward and backwards adjacency matrices, respectively: the two are adjacency matrices where the $N_\text{f}$ nodes with out-degree $k_\text{out}=0$ (for $\mathbf{A_\text{f}}$) and the $N_\text{b}$ nodes with zero in-degree (for $\mathbf{A_\text{b}}$) have been iteratively removed. Nodes with zero out-degree are removed from the forward process, and, analogously, nodes with zero in-degree are removed from the backward process, in order to ensure the ergodicity of both stochastic processes. With initial conditions $\mathbf{p}_\text{f}(0)=\frac{1}{N_\text{f}}\mathbf{1}$ and $\mathbf{p}_\text{b}(0)=\frac{1}{N_\text{b}}\mathbf{1}$ the covariance matrices of both processes are defined as:

\begin{equation}
    \mathbf{S}_\text{forw}(t)=\mathbf{P}_\text{f}(0)\mathbf{T}_\text{f}(t)\mathbf{T}_\text{f}^\text{inv}(t)-\mathbf{p}_\text{f}(0)^\textsf{T}\mathbf{p}_\text{f}(0)\\=\frac{1}{N_\text{f}}\mathbf{T}_\text{f}(t)\mathbf{T}_\text{f}^\text{inv}(t)-\frac{1}{N^2_\text{f}}\mathbf{\hat{1}}
\end{equation}
\begin{equation}
    \mathbf{S}_\text{back}(t)=\mathbf{P}_\text{b}(0)\mathbf{T}_\text{b}(t)\mathbf{T}_\text{b}^\text{inv}(t)-\mathbf{p}_\text{b}(0)^\textsf{T}\mathbf{p}_\text{b}(0)\\=\frac{1}{N_\text{b}}\mathbf{T}_\text{b}(t)\mathbf{T}_\text{b}^\text{inv}(t)-\frac{1}{N^2_\text{b}}\mathbf{\hat{1}}
\end{equation}
The covariance matrix of each process encodes the probability of two random walkers starting on two nodes $i$ and $j$ and ending up on the same node at time $t$ minus the same probability for two independent random walkers \cite{doi:10.1126/sciadv.abj3063,Bovet_telegram}. Clustering these two matrices in diagonal blocks provides the forward and backward partitions, which group together nodes in communities of nodes that act as similar “sources” for the forward covariance, or “sinks” for the backward covariance, of the flow. To find the optimal partitions, one can employ any clustering algorithm, such as the Louvain algorithm \cite{Blondel_2008}. Repeating the community detection for several iterations leads to an ensemble of possible partitions for the forward and backward covariance matrices. The forward and backward quality functions that are optimized to find the corresponding best forward or backward partition are given by the traces of the clustered forward and backward covariance matrices \cite{renaud_mauricio,Markov_st,doi:10.1126/sciadv.abj3063}. We recall how in this framework, the \emph{Markov time}, i.e. the time variable in each of the diffusion processes, represents the spatial resolution parameter of the community detection: a large variety of smaller communities are found for smaller Markov times, while bigger and less numerous communities are detected for larger Markov times. Thus partitions with optimal values of Flow Stability observed at different Markov times correspond to significant community structures of the network at different spatial scales.\\

In this work, we propose to adapt Flow Stability to static, weighted, directed networks with \emph{missing edges}. We do so by incorporating in the method an estimate of the experimental error $\mathbf{\epsilon}$, or uncertainty, on the ``measured'' out-strength of nodes, i.e., the number of missing out-going edges from each node. We consider a directed, weighted graph $G$, with adjacency matrix $A$, corresponding to the \emph{incomplete} or measured graph as sketched in Figure \ref{fig:schema}.a. We define the vector $\mathbf{\epsilon}$ as the 1-D vector of length $N$ whose $i$-th element $\epsilon_i$ represents the error on the out-strength of node $i$, $s_\text{out}^i$ (See schema in Figure \ref{fig:schema}.a for a visual explanation). We thus obtain the matrix $\mathbf{A}^\text{f}_{\Delta}$ iteratively removing from the original adjacency matrix all nodes with zero out-strength as well as zero error, thus only keeping the $N_{err}$ nodes with non-zero out-strength, or zero out-strength, but non-zero error. 
Similarly to regular FS, we obtain the adjacency matrix for the backward process, $\mathbf{A}_b$, iteratively removing from the network nodes with zero in-strength. We thus consider the two continuous time random walk processes, the forward process and the backward one. The transition matrix of the forward process is $\mathbf{T}_\Delta^\text{f}=e^{-t(\mathbf{I}-\mathbf{M}^\text{f})}$, where the matrix element of matrix $\mathbf{M}^\text{f}$ is defined as:

\begin{equation}
   M_{ij}^\text{f} = (1 - \alpha_i)\frac{A^\text{f}_{\Delta}(i,j)}{s^\text{out}_i}+\alpha_i\frac{s^\text{in}_j}{\sum_l s^\text{in}_l}
\end{equation}

Where 
\begin{align*}
s_\text{out}^i&\equiv\sum_j A^\text{f}_{\Delta}(i,j),\\
s_\text{in}^j&\equiv\sum_i A^\text{f}_{\Delta}(i,j),\\
\alpha_i&\equiv \frac{\epsilon_i}{\epsilon_i+s_\text{out}^i}.
    \end{align*}

Thus, in the forward process, we introduce a \emph{biased} teleportation \cite{biased,BRIN1998107,jure_sup_rand} term (Figures \ref{fig:schema}.b-c), corresponding to the second term in the r.h.s. of equation (3): a random walker on node $i$ is randomly teleported to another node with probability proportional to $\alpha_i$, and, in particular, it is teleported to node $j$ with a probability proportional to the in-strength $s^\text{in}_j$ of $j$. The total probability for the random walker on node $i$ to be teleported to node $j$ is thus $\alpha_i\frac{s^\text{in}_j}{\sum_l s^\text{in}_l}$. Such an approach is inspired by \cite{jelena_unweighted,Jelena_weighted}, where Smiljani\'c and collaborators use a Bayesian approach to reconstruct the community structure of an incomplete network with a maximum likelihood estimation of the transition matrix of a random walk. In \cite{Jelena_weighted}, the authors show how the obtained maximum likelihood estimator for the transition rates (matrix elements of the transition matrix) together with their prior distribution ``resemble modeling network flows with teleportation''. In fact, this approach is equivalent to that presented in \cite{Wang2008DirichletRankST}, where the authors propose the DirichletRank, a variant of PageRank \cite{BRIN1998107} adapted to solve the ``zero-one gap'' problem in PageRank. Such a solution is obtained by computing transition probabilities in the random surfing model using Bayesian estimation with a Dirichlet prior.
In our approach, this translates into incorporating the teleportation probability with a similar functional form as the one presented in the definition of DirichletRank \cite{Wang2008DirichletRankST}.
The transition matrix of the backward process is the same as in the regular FS: $\mathbf{T^\text{b}_{\Delta}}(t)=e^{-t\mathbf{L_\text{b}}}$, with $\mathbf{L_\text{b}}=\mathbf{I}-\mathbf{D}_\text{in}^{-1}\mathbf{A}_\text{b}^{T}$. With initial conditions $\mathbf{p}^\text{f}_{\Delta}(0)=\frac{1}{N^\text{f}_{\Delta}}\mathbf{1}$ and $\mathbf{p}^\text{b}_{\Delta}(0)=\frac{1}{N^\text{b}_{\Delta}}\mathbf{1}$ the covariance matrices of both processes are defined as:

\begin{equation}
\begin{split}
    \mathbf{S}^\text{f}_{\Delta}(t)& =\mathbf{P}_{\Delta}^\text{f}(0)\mathbf{T}_{\Delta}^\text{f}(t)\mathbf{T}_{\Delta}^\text{f,inv}(t)-\mathbf{p}_{\Delta}^\text{f}(0)^\textsf{T}\mathbf{p}_{\Delta}^\text{f}(0) ,\\ \mathbf{S}_{\Delta}^\text{b}(t)& =\mathbf{P}_{\Delta}^\text{b}(0)\mathbf{T}_{\Delta}^\text{b}(t)\mathbf{T}_{\Delta}^{b,inv}(t)-\mathbf{p}_{\Delta}^\text{b}(0)^\textsf{T}\mathbf{p}_{\Delta}^\text{b}(0)
    \end{split}
\end{equation}

Analogously to what is done in FS, we obtain two optimal partitions, one for the forward and one for the backward process. Nodes that are only in $\mathbf{A}^\text{f}_{\Delta}$ are clustered according to their forward communities, nodes that are only in $\mathbf{A}_\text{b}$ are clustered according to their backward communities and nodes that are in both networks are clustered according to the intersection of the forward and backward partitions. 
\section{Results}

\subsection{Synthetic Networks}

\begin{figure}[h!]
    \centering
    \includegraphics[width=1\textwidth]{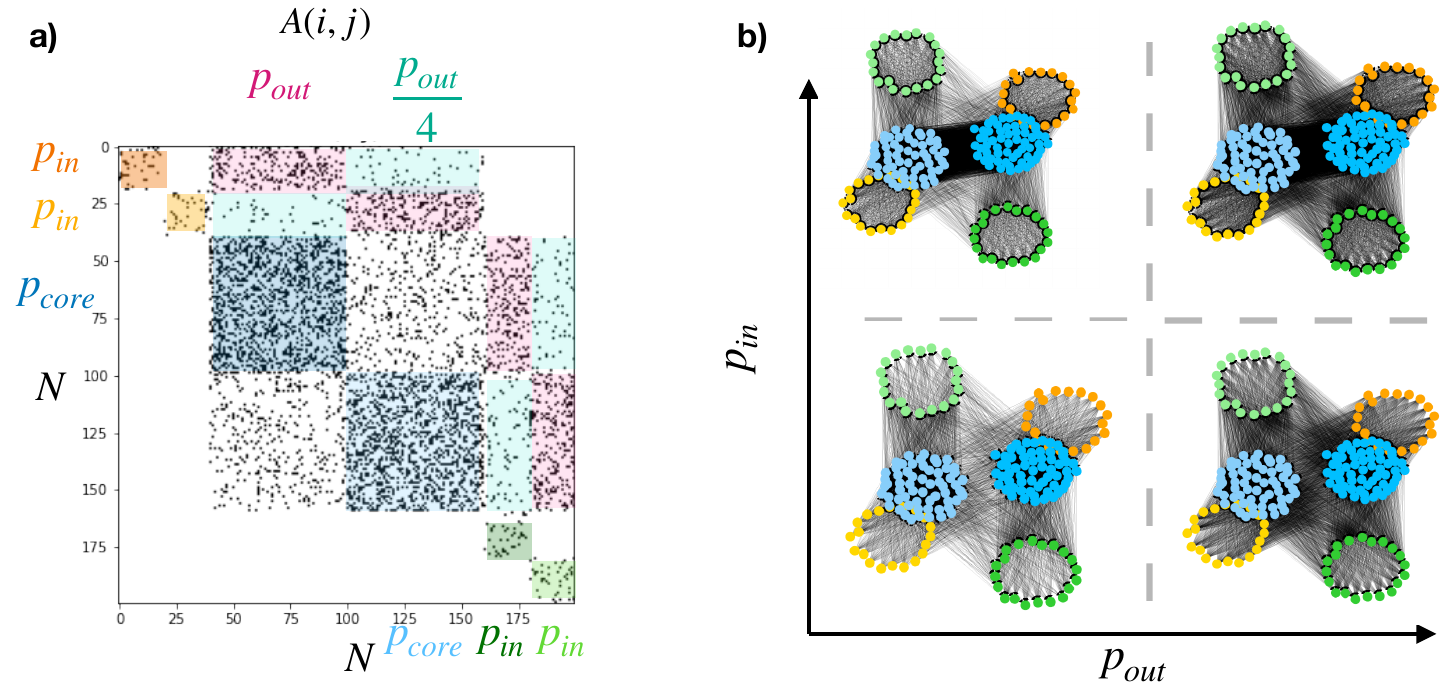}
    \caption{\textbf{Parameters space of the SBM ensemble:} \textbf{a)} example plot of the adjacency matrix of a network generated with the SBM where we highlight the two groups of source nodes (orange and yellow), whose within-connection probability is $p_{in}$ and that connect to their relative ``main" core with probability $p_\text{out}$, and ``secondary" core with probability $p_\text{out}/4$; the two cores of the network, internally interconnected with probability $p_\text{core}=0.4$, connected to each other with probability $p_\text{in}$, and each connected to their ``main" group of sink nodes (dark and light green) with probability $p_\text{out}$, and with probability $p_\text{out}/4$ to their secondary group of sinks. \textbf{b):} examples of the structure of the networks generated with different values of $p_\text{in}$ and $p_\text{out}$.}
    \label{fig:SBM-example}
\end{figure}

We first test the capability of $\Delta$ Flow Stability, compared with regular Flow Stability, to recover the original partition into communities of the observed network on synthetic networks.
In particular, we investigate the performance of $\Delta$FS on an ensemble of Stochastic Block Models. The structure of such networks corresponds to two sub-networks $G_1$ and $G_2$ made of a group of sources, a core, and a group of sinks. The generative model has two fixed parameters: the size of the network $N=200$, the connection probability within the two groups of core nodes, $p_\text{core}=0.4$. The free parameters are the connection probability between nodes belonging to the same group of source or sink nodes, $p_\text{in}$, which also corresponds to the probability of nodes in the two different cores to connect with each other; and the probability of each group of sources to be connected to their \textit{relative} core, $p_\text{out}$, which also corresponds to the probability of nodes in each of the two cores to connect to nodes in the respective groups of sinks. 
Therefore, the two initial sub-networks are coupled with each other using a nonzero probability of source nodes of one sub-network to connect to core nodes of the other sub-network $p^{\text{source},G_1}_{\text{core},G_2}=p^{\text{source},{G_2}}_{\text{core},G_1}=\frac{p_\text{out}}{4}$. Core nodes of $G_1$ and $G_2$ also have a nonzero probability of being connected to nodes in the group of sinks of the other sub-network. The network under study thus corresponds to a \emph{double} weighted, directed core-periphery structure. An example of one adjacency matrix generated with such an approach is in Figure \ref{fig:SBM-example}a, while in Figure \ref{fig:SBM-example}.b we display a visual explanation of the impact of the two free parameters $p_\text{out}$ and $p_\text{in}$ on the network structure.

We generate SBMs by varying the free parameters $p_\text{in}\in[0,0.35]$ and $p_\text{out}\in[0.1,p_\text{core}]$. For each pair of values of $(p_\text{in},p_\text{out}$ we perform the following experiment.

We iteratively remove an increasing fraction of randomly selected edges from the network. For each value of the fraction $r$ of randomly removed edges, we detect the community structure of the ``new" network via Flow Stability, and with $\Delta$FS, at each Markov time (resolution parameter), we perform 50 runs of the Louvain algorithm, and select the partitions maximizing FS and $\Delta$ FS. We quantify the performance in recovering the community structure of the original network of the two methods by computing the Normalized Mutual Information (\text{NMI}) between each of the two partitions obtained by the two methods, and the original, $6$-blocks, partition. 

\begin{figure}[h!]
    \centering
    \includegraphics[width=\textwidth]{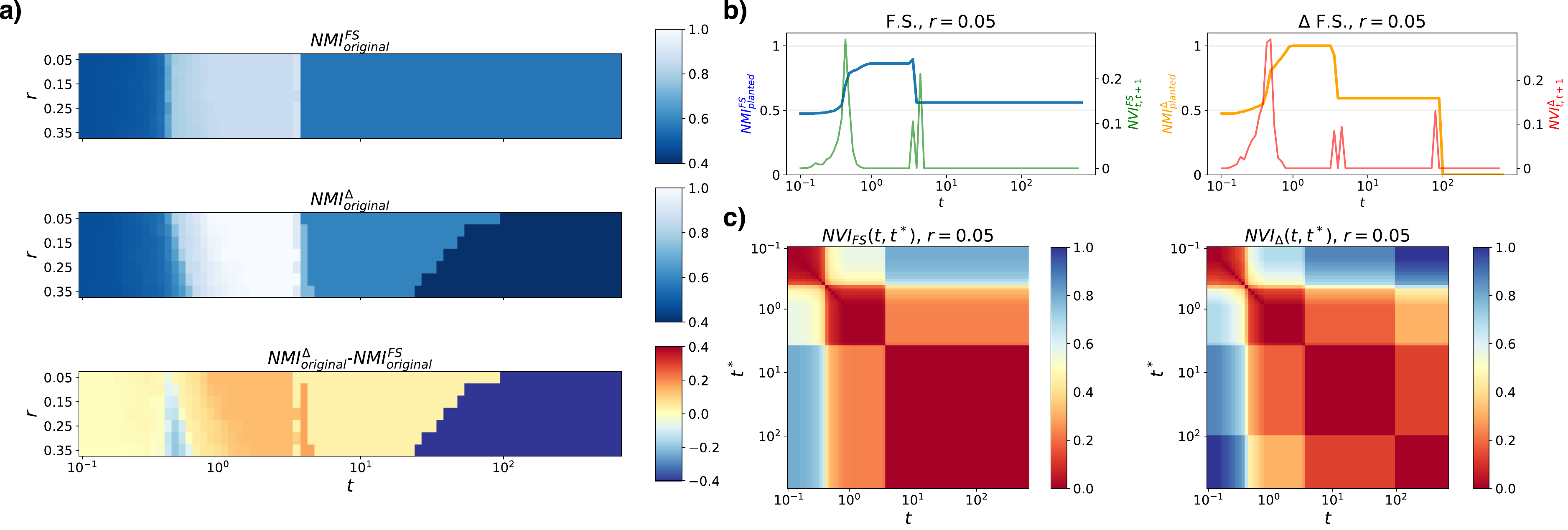}
    \caption{\textbf{Recovering community structure on a synthetic network:} \textbf{a)} (top) heatmap of the Normalized Mutual Information computed between the Flow Stability partition at specific values of Markov time $t$ ($x$-axis) and fraction of randomly removed edges $r$ ($y$-axis) and the original partition in the SBM; (middle) heatmap of the Normalized Mutual Information computed between the $\Delta$ Flow Stability partition at specific values of Markov time $t$ ($x$-axis) and fraction of randomly removed edges $r$ ($y$-axis) and the original partition in the SBM; (bottom) heatmap of the difference of the previous two heatmaps. \textbf{b):} (left) plot of the $\text{NMI}$ computed at each value of the Markov time $t$, between the partition obtained by Flow Stability with $r=0.05$ and the original partition of the SBM (left y-axis, in blue) and plot of the NVI computed between the partitions obtained via FS at two consecutive Markov times $t$ and $t+1$ (right y-axis, in green); (right) plot of the $\text{NMI}$ computed at each value of the Markov time $t$, between the partition obtained by $\Delta$FS with $r=0.05$ and the original partition of the SBM (left y-axis, in orange) and plot of the NVI computed between the partitions obtained via $\Delta$FS at two consecutive Markov times $t$ and $t+1$ (right y-axis, in red). \textbf{c):}  heatmap of \text{NVI} computed between each pair of partitions obtained with FS (left), or $\Delta$FS (right), at different Markov times $t$ and $t^*$, both for $r=0.05$.}
    \label{fig:SBM complex}
    \end{figure}

An example of the results obtained for such analysis on one initial network sampled from the generative model with $p_\text{in}=0.1$, $p_\text{out}=0.2$, is shown in Figure \ref{fig:SBM complex}. In Figure \ref{fig:SBM complex}.a, where we plot the heatmap of \text{NMI} between the each of the partitions obtained with FS (top) and $\Delta$FS (middle) and the original partition, for each value of $r$ and each Markov time $t$, we can observe how already for $r=0.05$, the original partition cannot be recovered exactly through Flow Stability, while even with $r=0.35$, thus with $35\%$ of edges randomly removed, we find a time scale (values of the Markov time) for which $\Delta$FS recovers the original community structure of the network. 

In Figure \ref{fig:SBM complex}.b we plot the value of the NMI between the partition obtained by FS (left, in blue) and with $\Delta$FS (right, in orange) at different values of the respective Markov times for a fixed value of the fraction of randomly removed edges: $r=0.05$. Here we note how, for some Markov times, $\Delta$FS recovers exactly the original partition, while the NMI between the partition obtained with FS and the original partition never attains 1. In the same figures, we plot as a function of the Markov time, the Normalized Variation of Information (NVI)  \cite{doi:10.1126/sciadv.abj3063,Bovet_telegram} between the partitions obtained by FS at two consecutive Markov times $t$ and $t+1$ (left, in green), and, analogously, the NVI between the partitions obtained by $\Delta$FS (left, in green) at two consecutive Markov times (right, in red). The NVI between two partitions is zero when they are identical. The more different the two partitions are, the closer their NVI is to 1. Peaks in these two curves thus highlight that the resolution parameter, the Markov time, allows the detection of the presence of different \emph{scales} in the community structure of the network under study. Furthermore, we note how the plateaus in the curves obtained from the comparison to the original partition of each partition obtained with the two methods are comprised of two peaks of the green (left, for FS) and red (right, for $\Delta$FS) curves. The presence of several scales of the community structure of the network is evidenced in Figure \ref{fig:SBM complex}.c, where we plot two colormaps, one for FS (left, blue) and one for $\Delta$FS (right, orange). Entries of each colormap correspond to the value of NVI computed between the partitions obtained with one of the two methods at any two Markov times $t$ and $t^*$. Diagonal blocks of zero NVI in these two colormaps suggest \emph{periods} of Markov times where the community structure is unchanged, as detected by one of the two methods. The number of diagonal blocks thus corresponds to the number of spatial scales of the network's community structure that each method can capture \cite{Markov_st,lambiotte2021modularity}. We note in Figure \ref{fig:SBM complex}.c, on the left, how for FS we can clearly observe three diagonal blocks of zero NVI and thus three scales, also clearly separated by the two peaks in the green curve of $\text{NVI}^{FS}(t,t+1)$ in Figure \ref{fig:SBM complex}.b, on the left. For $\Delta$FS in Figure \ref{fig:SBM complex}.c, on the right, we observe a more complex pattern: both for low values of Markov times as well as for values of the Markov time above one, we observe two clear diagonal blocks, respectively identified by two peaks of $\text{NVI}^{\Delta}(t,t+1)$ in Figure \ref{fig:SBM complex}.b; Instead, for intermediate values of the Markov time, we observe more than one diagonal block. The first of such blocks corresponds to the range of Markov times in which $\Delta$FS recovers the original partition exactly. In each of the following short intervals of Markov time values, the accordance with the original partition gradually decreases, suggesting how, while $\Delta$FS might be capturing finer changes in the spatial scale of the community structure of the network, the partition that it is retrieving is further away from the original community assignments of nodes.

\begin{figure}[h!]
    \centering
    \includegraphics[width=\textwidth]{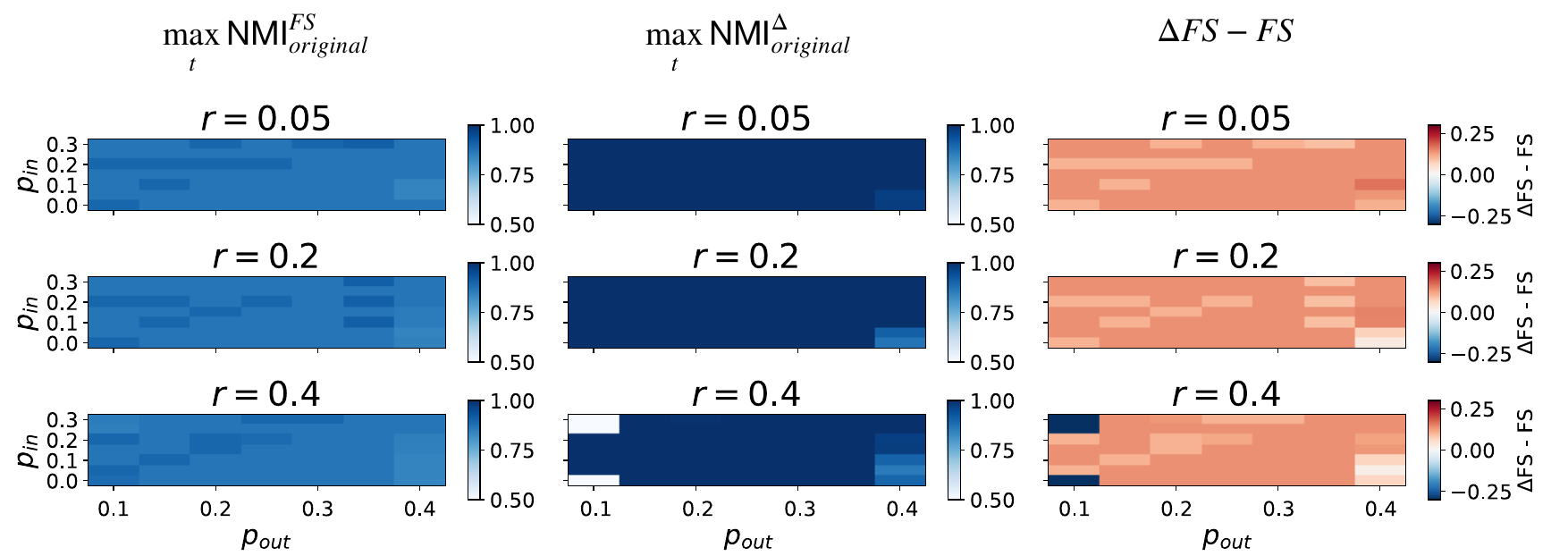}
    \caption{\textbf{Recovering the community structure of an ensemble of stochastic block models:} Left - heatmap of the maximum value of the NMI between the partition obtained with FS and the original partition at each Markov time for each pair of values of $(p_\text{in},p_\text{out})$ for several values of the fraction of randomly removed edges $r$. Middle - heatmap of the maximum value of the NMI between the partition obtained with $\Delta$FS and the original partition at each Markov time for each pair of values of $(p_\text{in},p_\text{out})$ for several values of the fraction of randomly removed edges $r$. Right - difference between the left and middle columns.}
    \label{fig:SBM-pin-pout}
\end{figure}

In Figure \ref{fig:SBM-pin-pout} we plot, for each pair $(p_\text{in},p_{out})$, the maximum value of the \text{NMI} between the partition obtained with FS, $\max_t \text{\text{NMI}}^\text{FS}_{original}$ (left), or with $\Delta$FS, $\max_t \text{\text{NMI}}^{\Delta}_{original}$ (middle), and the original partition over the Markov time of the random walks, for several values of the fraction of randomly removed edges $r$. On the right of Figure \ref{fig:SBM-pin-pout}, we plot the difference between the two heatmaps. We note how for low values of $p_\text{out}$, $\Delta$ Flow Stability can recover the original partitions only for fractions of randomly removed edges up to $r=0.4$. For higher values of $r$, the teleportation in the random walk is the reason for a higher probability of overfitting in $\Delta$FS, with respect to FS. On the other hand, both methods fail for large values of $r$ when $p_\text{out}=p_\text{core}$, as in both cases, the 2-block partition becomes more likely for both methods.

\subsection{Telegram Network}
Telegram is a messaging platform that allows users to create and join public group chats and broadcasting channels, where posts and messages can contain mentions or URL links to other group chats and channels on the platform. In this work, we analyze the dataset first presented in \cite{Bovet_telegram}.
The dataset contains existing messages from public Telegram channels/groups related to the UK far-right. As Telegram users are free to delete their messages, the ensemble of collected messages does not include previously deleted messages. We build a directed weighted network where nodes represent Telegram channels/groups and where an edge from node A to node B can represent a mention of node B in node A, a URL link pointing to node B posted in node A, or a message from node B forwarded in node A.

\subsubsection{Community detection with Flow Stability}

\begin{figure}[h!]
    \centering\includegraphics[width=0.6\linewidth]{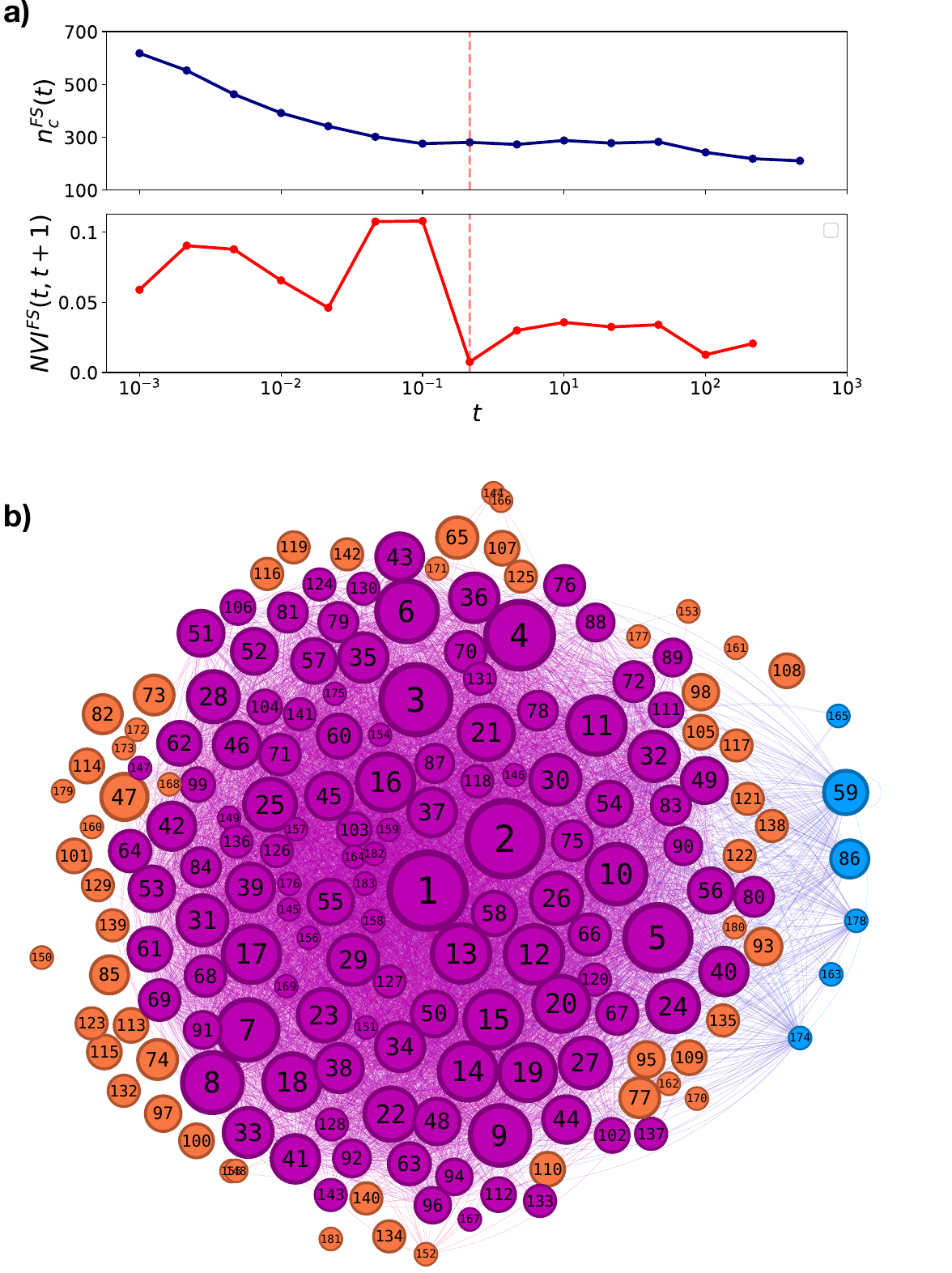}
    \caption{\textbf{Flow stability clustering of the Telegram channels network.}\textbf{a):} the curves in the plot correspond to the number of clusters $n_c^\Delta(t)$ (top, in dark blue) found with $\Delta$FS at each Markov time, t; the Normalized Variation of Information between two partitions obtained at adjacent Markov times with $\Delta$FS, $\text{NVI}^\Delta(t,t+1)$ (bottom, in red).  \textbf{b)} Nodes represent the communities, and edges represent the number of links between channels in each community. Upstream communities are shown in blue. Core communities are in brown, and downstream communities in purple. The labels indicate the rank of each community in terms of their size.}
\label{fig:flow_stab_tel_2}
\end{figure}

In \cite{Bovet_telegram}, the authors separate the dataset into three networks, each corresponding to a different period of the evolution of the overall system. In this work, we treat it as a single network. We thus employ Flow Stability to investigate the community structure of such a network (Figure \ref{fig:flow_stab_tel_2}). Due to the size of the network, which includes $N=12,653$ nodes and $E=6,925,865$ edges, we adopt the linear approximation of the transition matrices as presented in \cite{doi:10.1126/sciadv.abj3063}, and further employ a similar approach for $\Delta$ Flow Stability (see Appendix A).
Via regular Flow Stability, we retrieve an optimal partition of the network corresponding to a value of the Markov time $t$, the resolution parameter in Flow Stability, of $t=8.86\cdot10^{-1}$ (highlighted by the vertical dotted line in Figure \ref{fig:flow_stab_tel_2}.a).
The optimal partition is defined as the one corresponding to the minimum value of the Normalized Variation of Information computed between pairs of partitions obtained via FS at contiguous Markov times, $\text{NVI}(t,t+1)^{FS}$.
The optimal partition, which we refer to as $\Pi^o_\text{FS}$, is characterized by $n_c=183$ separate clusters. In continuity with \cite{Bovet_telegram}, we investigate the functional role of each cluster by computing the in-balance $I_c$ between the number of edges connecting the nodes in a cluster $c$ to other nodes in the same cluster and the number of edges connecting nodes in $c$ to the rest of the network. We then define clusters with values of $I_c\leq0.2$ as up-stream clusters (in blue in Figure \ref{fig:flow_stab_tel_2}.b), clusters with values of $I_c\geq.8$ as down-stream clusters (in orange in Figure \ref{fig:flow_stab_tel_2}.b) and clusters with values of $0.2<I_c<0.8$ as core clusters (in purple in Figure \ref{fig:flow_stab_tel_2}.b). We observe how $\Pi^o_\text{FS}$ is characterized by $122$ core clusters, a vast majority compared to the 55 down-stream clusters and the six up-stream clusters that are found. In Figure \ref{fig:flow_stab_tel_2}.b, where each node in the network represents a cluster in $\Pi^o_\text{FS}$ color-coded according to its function, the size of nodes is proportional to the number of Telegram channels in that cluster, i.e., the size of the cluster. We note how core clusters are generally larger. In Table \ref{table:flow_stab_tel_2}, we list the size rankings of the two largest clusters of each different functional role. We observe that the largest downstream cluster is only 47th (second largest is 65th) in the size-wise ranking of the clusters, while the largest upstream cluster is 59th (second largest is 86th).
\begin{table}[h!]
    \centering
    \begin{tabular}{|c|c|}
    \toprule
      Top two clusters   & Size Rank \\
      \midrule
      Core &1st \& 2nd\\
       \midrule
       Downstream    & 47th \& 65th \\
       \midrule
       Upstream    & 59th \& 86th \\
       \bottomrule
    \end{tabular}
    \caption{Ranking by size of the two largest clusters of the three types, i.e., core, downstream and upstream, found in $\Pi^0_{\text{FS}}$.}
    \label{table:flow_stab_tel_2}
\end{table}


We perform two different analyses, employing $\Delta$ Flow Stability, on the dataset that we describe above: 
First, we repeat the analysis presented in Section 3.1, thus removing an increasing fraction of randomly selected edges from the original Telegram network. We use the information on removed edges as an estimate of the experimental error on the \emph{measured} out-strength of nodes and feed it to our $\Delta$ Flow Stability method. We consider $\Pi^o_\text{FS}$ as the \emph{ground truth} partition and compare how similar to $Pi^0_\text{FS}$ are the partition obtained by FS and $\Delta$FS for every fraction of randomly removed edges.
In our second analysis, we directly estimate the amount of deleted out-links per Telegram channel from the data and use it in $\Delta$FS, thus finding a community structure of the network adjusted to false negatives in the links. We compare this result to $\Pi^0_\text{FS}$ and discuss the main differences/similarities between the two partitions.

\subsubsection{Community structure recovery after random edge removal in the Telegram dataset}
We first perform the same analysis on the Telegram dataset as on the ensemble of SBM-generated networks presented in Section 3.1. We first define as \textit{original} partition of the network $\Pi^0_\text{FS}$. We then iteratively remove an increasing fraction of randomly selected edges from the network, starting with $2.5\%$, up to $25\%$. For each value of the fraction $r$ of randomly remove edges, we detect the community structure of the new network via FS and with $\Delta$FS, at each Markov time. We perform 50 runs of the Louvain algorithm and select the partition maximizing the Flow or $\Delta$ Flow Stability. We quantify the performance in recovering the community structure of the original network of the two methods by computing the Normalized Mutual Information (\text{NMI}) between each of the two partitions obtained by the two methods and the original partition. 

\begin{figure}[h!]
   \centering\includegraphics[width=\textwidth]{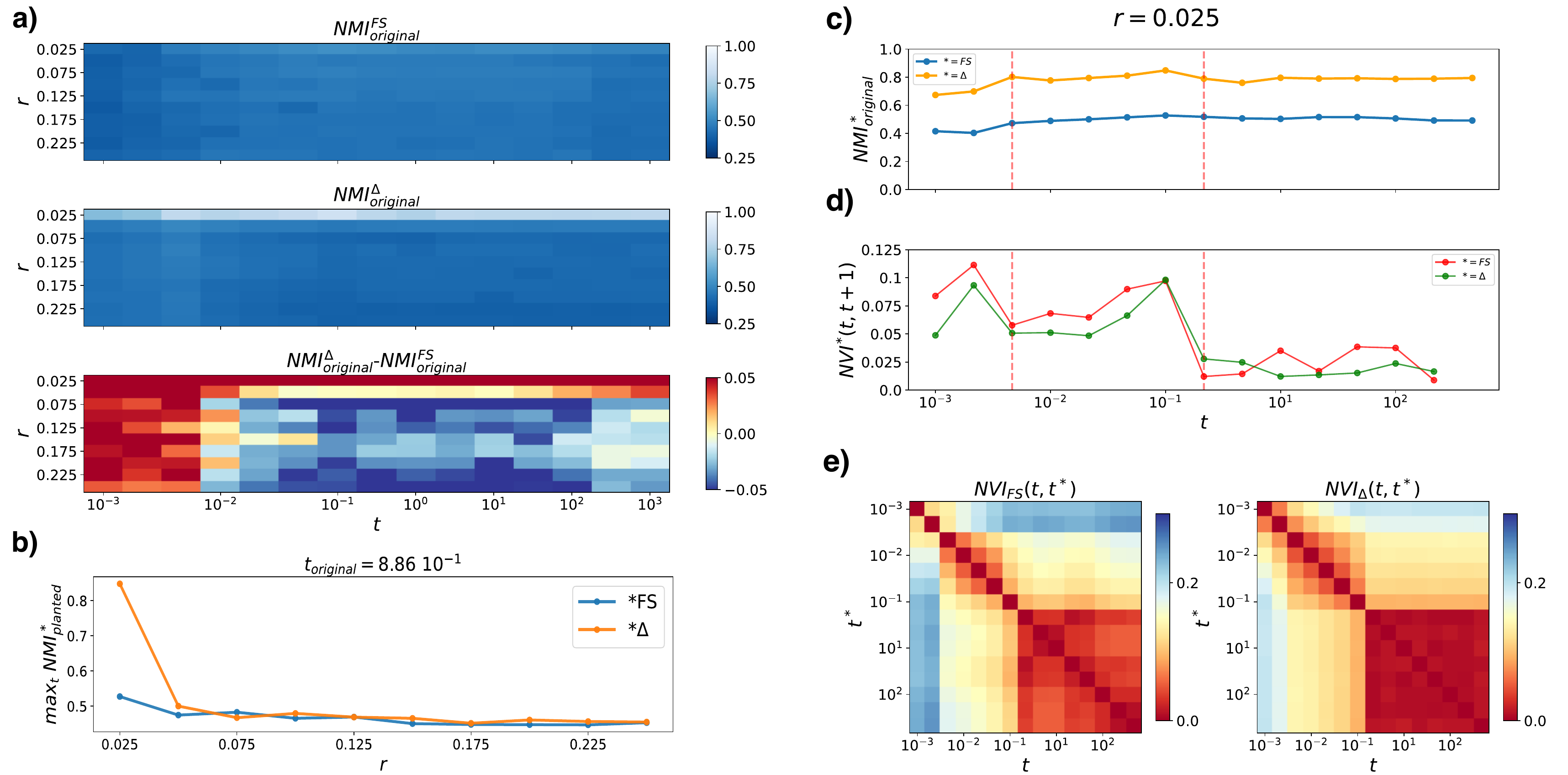}
    \caption{\textbf{Community structure recovery after random edge removal in the Telegram dataset:} \textbf{(a)} (top) heatmap of the $\text{NMI}$ computed between the FS partition at specific values of Markov time $t$ ($x$-axis) and fraction of randomly removed edges $r$ ($y$-axis) and $\Pi^0_\text{FS}$, i.e.,the original partition in the Telegram network; (middle) heatmap of the $\text{NMI}$ computed between the $\Delta$FS partition at specific values of Markov time $t$ ($x$-axis) and fraction of randomly removed edges $r$ ($y$-axis) and $\Pi^0_\text{FS}$; (bottom) heatmap of the difference of the previous two heatmaps. \textbf{b):} plot of the maximum over the Markov time $t$ of the $\text{NMI}$ computed at each value of the fraction of randomly removed edges $r$, between the partition obtained by FS and $\Pi^0_\text{FS}$ (in blue), and between the partition obtained by $\Delta$FS and $\Pi^0_\text{FS}$, in orange. \textbf{c):} in the case where $r=0.025$, and thus the $2.5\%$ of the edges in the network have been randomly removed, we plot the $\text{NMI}$ computed between $\Pi^0_\text{FS}$ and the partition obtained with FS (in blue) and $\Delta$FS in orange as a function of the Markov time $t$. \textbf{d)}:  Normalized Variation of Information between the partitions obtained at each pair of consecutive Markov times $\text{NVI}(t_k,t_{k+1})$ with FS (in red) and with $\Delta$FS (in green). \textbf{e):} (left) heatmap of \text{NVI} computed between each pair of partitions obtained with FS at different Markov times $t$ and $t^*$; (right) heatmap of \text{NVI} computed between each pair of partitions obtained with $\Delta$FS at different Markov times $t$ and $t^*$. In both cases, $r=0.025$ }
   \label{fig:teleg_exp_1}
\end{figure}

The results are shown in Figure \ref{fig:teleg_exp_1}. In Figure \ref{fig:teleg_exp_1}.a, we plot the heatmap of \text{NMI} between each of the partitions obtained with FS (top) and $\Delta$FS (middle) and the original partition, for each value of $r$ and each Markov time $t$, we can observe how $\Delta$FS recovers partitions closer to the original one at lower Markov times: this is highlighted in the bottom heatmap, where we plot the difference between the two previous heatmaps. To assess which of the two methods recovers the most similar partition in comparison to the original community structure of the network, in Figure \ref{fig:teleg_exp_1}.b, we plot the maximum value over the Markov time $t$ of the Normalized Mutual Information computed between the original partition and that obtained with FS (in blue) or with $\Delta$FS (in orange) for each fraction of randomly removed edges $r$. Here we see how for $r=0.025$, after removing $2.5\%$ of the edges in the original network, $\Delta$FS detects an underlying community structure which has values of NMI of up to $0.9$ w.r.t. the original partition $\Pi^0_\text{FS}$, well above the result obtained via regular FS. As $r$ increases, the performance of $\Delta$ Flow Stability also drops sensibly: already for $r=0.05$, we note a dramatic decrease for $\Delta$FS, which achieves performances comparable to FS. An inspection of the results shows that this is due to the incorrect partitioning of nodes that, in the original partition, are in the two largest clusters (see Table \ref{table:flow_stab_tel_2}). They are divided into smaller clusters by both FS and $\Delta$FS for $r>0.025$. For higher values of $r$, thus, the optimal partition for $\Delta$FS is in lower accordance with $\Pi^0_\text{FS}$ in comparison to partitions found at lower Markov times, highlighted by the red area for low Markov times in \ref{fig:teleg_exp_1}a, bottom. We verify this by computing the NMI between $\Pi^0_\text{FS}$ and the partition obtained via $\Delta$FS with $r=0.025$ and $r=0.05$ only taking into account the nodes in the network that are not included the second largest cluster of $\Pi^0_\text{FS}$, obtaining values of $\text{NMI}(\Pi^0_\text{FS};\Delta \text{FS},r=0.25)=0.96$ and $\text{NMI}(\Pi^0_\text{FS};\Delta \text{FS},r=0.5)=0.65$. We thus repeat the analysis, this time disregarding all nodes included in the two largest clusters in $\Pi^0_\text{FS}$, that, combined, include roughly $\approx 23\%$ of the nodes in the network. In this case, the accordance between $\Pi^0_\text{FS}$ and the partitions found via $\Delta$FS increases even further, with $\text{NMI}(\Pi^0_\text{FS};\Delta \text{FS},r=0.25)=0.97$ and $\text{NMI}(\Pi^0_\text{FS};\Delta \text{FS},r=0.5)=0.85$, hence, much higher than what is found when the nodes in the two largest clusters in $\Pi^0_\text{FS}$ are included, as in Figure \ref{fig:teleg_exp_1}.b.

We compare the results obtained by FS and $\Delta$FS with $r=0.025$ for several values of the Markov time in Fig. \ref{fig:teleg_exp_1}c.
In Figure \ref{fig:teleg_exp_1}.d, we plot the curves of $\text{NVI}$ computed between the partitions obtained with FS (red) and $\Delta$FS (green) between two consecutive Markov times. We highlight how both FS and $\Delta$FS capture the existence of at least two relevant scales of the community structure of the network, corresponding to the transient local minima of the $\text{NVI}$. While the optimal partitions for each method correspond to the partition found at $t=8.86\cdot 10^{-1}$, the partitions obtained for $t>8.86\cdot 10^{-1}$ capture the community structure of the underlying network at a coarser scale. We show evidence of the existence of multiple description scales of this network in Figure \ref{fig:teleg_exp_1}.e. Here we plot the heatmaps of NVI computed between two partitions obtained by FS (left) or $\Delta$FS (right) at any two Markov times $t,t^*$: while it is harder to visually detect the presence of three diagonal blocks in this matrices, the out-of-diagonal elements help us identify at least three different scales, corresponding to the sets of values of the Markov time, the resolution parameter, identified by the two optimal values highlighted in Figures \ref{fig:teleg_exp_1}.c-d.

\newpage
\subsubsection{Reconstruction of Telegram Network}
\begin{figure}[h!]
    \centering
    \includegraphics[width=0.7\textwidth]{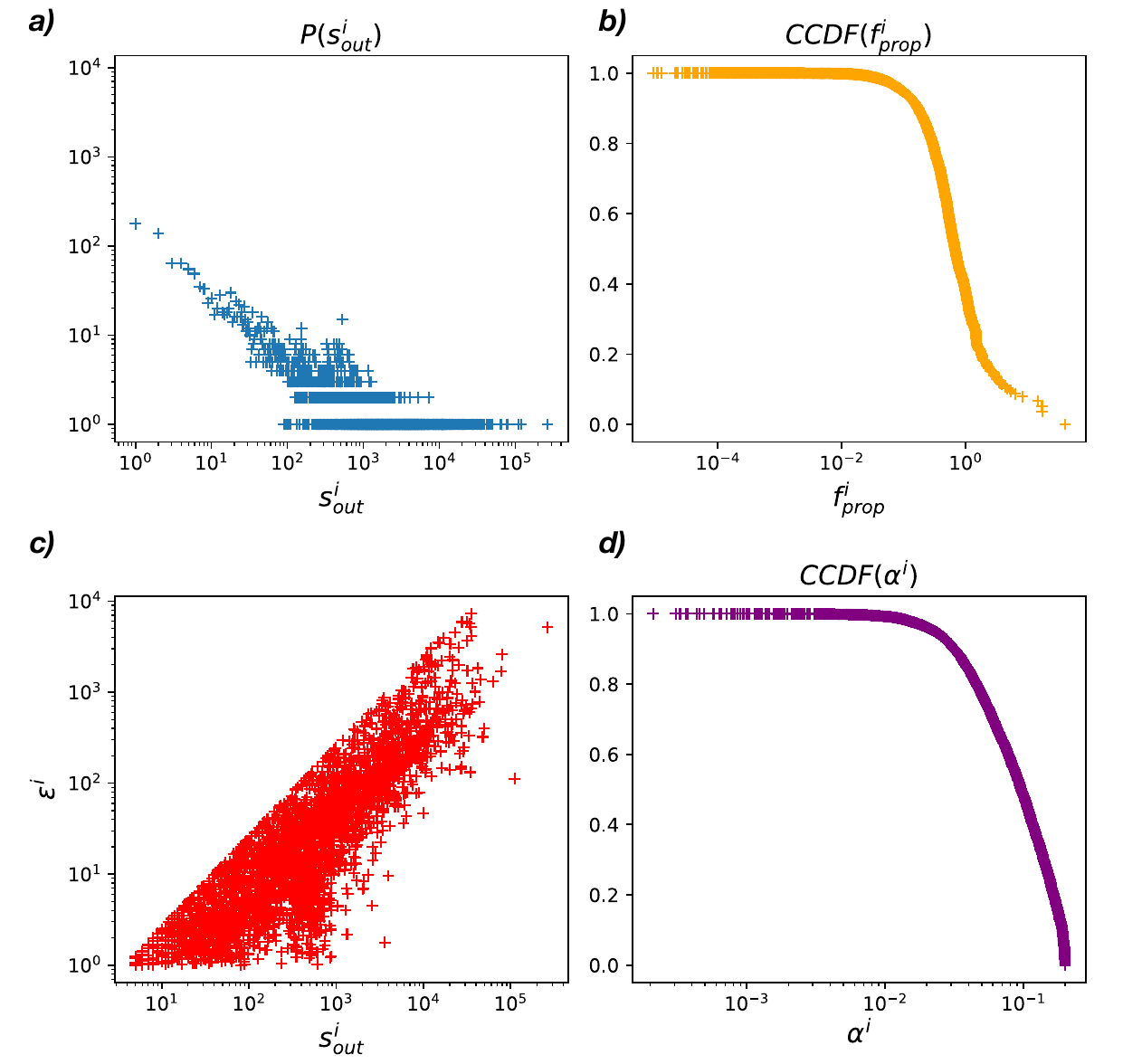}
    \caption{\textbf{Estimation of the error on out-strengths:} \textbf{a)} distribution of the out-strength $s_\text{out}$ of nodes. \textbf{b)} complementary cumulative distribution of the observed proportion of edges containing links to other channels $f_{prop}$ for each node. \textbf{c)} scatterplot of the out strength of nodes ($s^i_\text{out}$, x-axis)) and the relative experimental error $\epsilon^i$ on the out-strength of nodes (y-axis). \textbf{d)} complementary cumulative distribution of the teleportation probability $\alpha$.}
    \label{fig:error-esrimate}
\end{figure}

In the Telegram dataset, the out-strength $s_\text{out}^i$ of a channel (node in the network) $i$ corresponds to all the posts in that channel containing a mention to another channel i.e., a link to another node in the network. We plot the distribution of nodes' out-strength in Figure \ref{fig:error-esrimate}.a, in log-log scale, highlighting how such distribution is heavy-tailed. For any node $i$ in the network, we then define the \emph{observed proportion} $f^i_{prop}$ as the fraction of messages posted in that channel containing a mention to another channel, and thus representing a link in the network (see the distribution of $f^i_{prop}$ in Figure \ref{fig:error-esrimate}.b). In this dataset, we can access the number of deleted posts in a channel. We thus estimate the number of missing out-going edges from a node $i$, i.e., the error $\epsilon_i$, to correspond to the product of the number of deleted posts of node $i$ and $f_{prop}^i$. This leads to a distribution of nodes' experimental error on their out-strength $\epsilon_i$, which is also heavy-tailed. Furthermore, as can be observed in the scatter-plot in Figure \ref{fig:error-esrimate}.c, $\epsilon_i$ and $s_\text{out}^i$ are positively correlated variables, meaning that the higher the out-strength of a node in the network, the higher the experimental error on its out-strength. We then compute the teleportation probabilities $\alpha_i$ as described in Section \textbf{2}, and plot the distribution of $\alpha_i$ in Figure \ref{fig:error-esrimate}.d.

\begin{figure}[h!]
    \centering\includegraphics[width=0.6\linewidth]{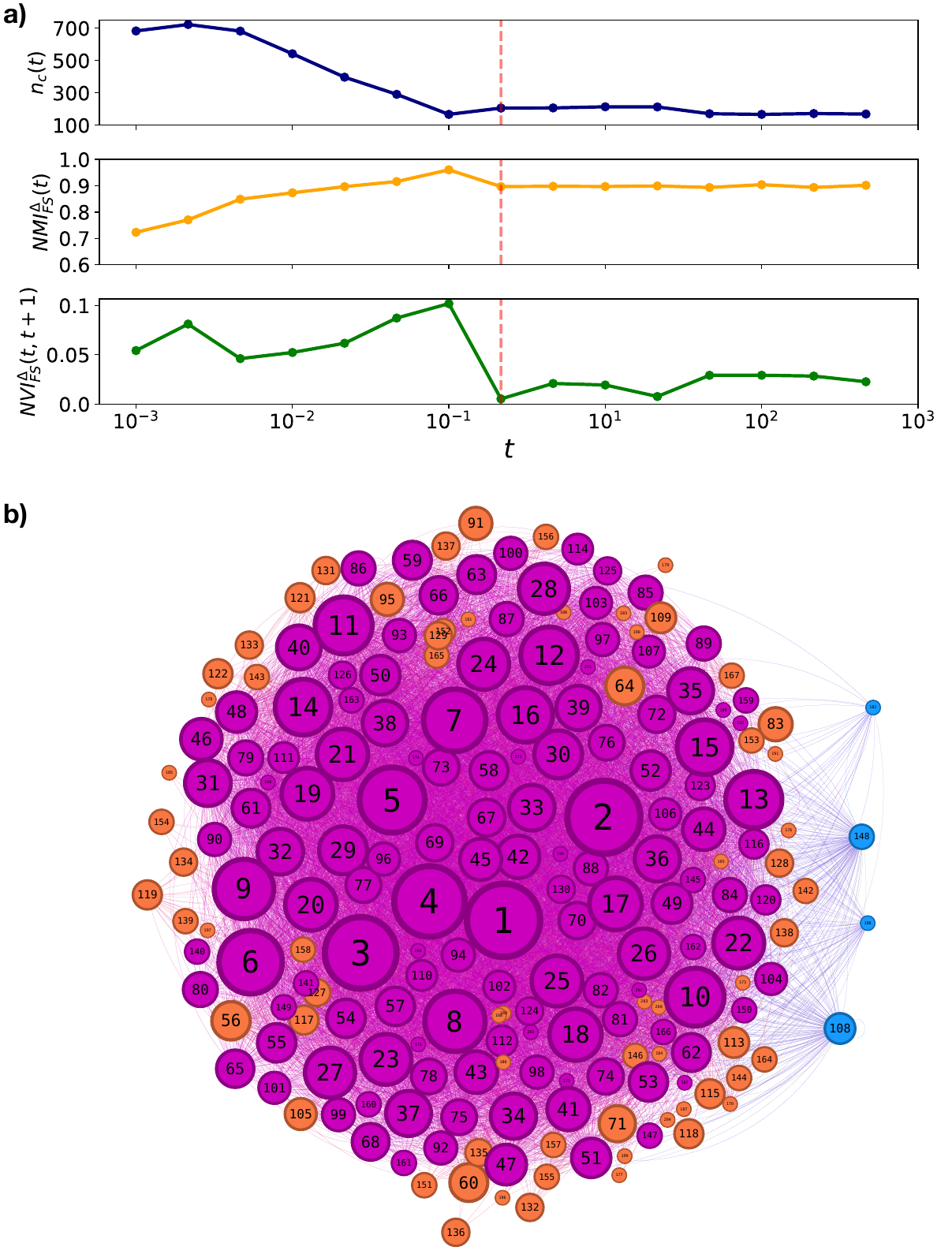}\caption{\textbf{Reconstruction of the community structure with $\Delta$ Flow Stability:} \textbf{a):}  number of clusters $n_c^\Delta(t)$ (top, in dark blue) found with $\Delta$FS at each Markov time (t); $\text{NMI}_\text{FS}^{\Delta}(t)$ computed between the partition obtained with $\Delta$FS at each Markov time and the optimal partition obtained with FS, $\Pi^0_\text{FS}$ (middle, in orange); the \text{NVI} between two partitions obtained at adjacent Markov times with $\Delta$FS, $\text{NVI}^\Delta(t,t+1)$ (bottom, in green). \textbf{b)} $\Delta$ Flow Stability clustering of the Telegram channels network. Nodes represent the communities and edges represent the number of links between channels in each community. Upstream communities are shown in blue, core communities in brown and downstream communities in purple. The labels indicate the rank of each community in terms of size and the size.}\label{fig:teleg_exp_2_error}
\end{figure}

We can thus inform $\Delta$ Flow Stability with the sequence of node-wise errors $\epsilon_i$, find an optimal network partition, and compare it to $\Pi^0_\text{FS}$. We plot the result of our analysis in Figure \ref{fig:teleg_exp_2_error}. Here we note how the optimal partition for $\Delta$FS is found at the same value of the Markov time as for $\Pi^0_\text{FS}$, i.e., $t=8.86\cdot 10^{-1}$ (Figure \ref{fig:teleg_exp_2_error}.a). The community structure of the overall network found with $\Delta$FS (Figure \ref{fig:teleg_exp_2_error}.b) is somewhat similar to $\Pi^0_\text{FS}$, which is reflected by the value of the $\text{NMI}$ w.r.t. the original partition, which is $\simeq 0.9$.
However, the two partitions are not identical. We thus investigate, in particular, the composition of the three largest clusters in the two partitions, measuring the Ranking-Biased Overlap (RBO) \cite{RBO_10.1145/1852102.1852106} between the lists of channels ranked by their Katz Centrality comprised in each of the clusters. The RBO is a similarity measure devised for comparing ranked lists, and it is designed to weigh high ranks more heavily than low. It thus assigns higher values to ranked lists with the same elements (Telegram channels in the clusters) with high ranking rather than the same elements of low ranking. In Table \ref{tab:RBO}, we note how the two largest clusters in the two partitions, $C^1_{\Pi^0_\text{FS}}$ and $C^1_{\Delta}$ are nearly identical, with an $\text{RBO}=\simeq0.999$, as well as the third largest clusters in the two partitions, namely $C^3_{\Pi^0_\text{FS}}$ and $C^3_{\Delta}$, with an $\text{RBO}=\simeq0.96$; however the second largest clusters in the two partitions, $C^2_{\Pi^0_\text{FS}}$ and $C^2_{\Delta}$, are far from being in complete accordance, suggesting that, nonetheless the community structure of this Telegram network, reconstructed taking into account the sampling of the edges, does not entirely coincide with what found with FS on the empirically constructed network.

\begin{table}[h!]
    \centering
    \begin{tabular}{|c|c|c|}
\toprule      $RBO(C^1_{\Pi^0},C^1_{\Delta})$   & $RBO(C^1_{\Pi^0},C^2_{\Delta})$ &$RBO(C^3_{\Pi^0},C^3_{\Delta})$ \\
\midrule
  0.999  &  0.683   & 0.963\\
  \bottomrule
    \end{tabular}
    \vspace{0.1cm}
    \caption{Rank-Biased Overlap (RBO), as defined in  \cite{RBO_10.1145/1852102.1852106}, computed with weight-decline parameter $p=0.9$ between each the top three clusters found in $\Pi^0$ ($C^1_{\Pi^0}$,$C^2_{\Pi^0}$,$C^3_{\Pi^0}$), and the corresponding top three clusters found via $\Delta$FS ($C^1_{\Delta}$,$C^2_{\Delta}$,$C^3_{\Delta}$) accounting for the uncertainty on nodes' out-strengths.}
    \label{tab:RBO}
\end{table}

\newpage
\section{Conclusion}

Unreliable network data can cause community-detection methods to highlight spurious structures with misleading information about the organization and function of complex systems \cite{newman_rich_noisy,Newman_bayesian,peixoto_reconstruct}.
In this work, We have presented $\Delta$ Flow Stability ($\Delta$FS), a modification to the Flow Stability (FS) community detection framework \cite{doi:10.1126/sciadv.abj3063}, equipping the method with a regulatory mechanism to deal with missing link observations in weighted and directed networks. By including a teleportation term in the transition matrix of the forward diffusive process, the network flow dynamics account for the uncertainty on observed node out-strengths. 
We tested the performance of $\Delta$FS, in comparison with FS, in retrieving the community structure of an ensemble of synthetic networks after removing an increasing fraction of randomly selected edges. We show how $\Delta$FS outperforms FS, thus proving more resilient to false negatives in link observations.

We analysed a real network composed of Telegram channels/groups, where users can delete their own messages, but where an estimate of the amount of deleted messages per channel/group can be found. First, we perform Flow Stability on the original dataset, following the approach presented in \cite{Bovet_telegram}, thus obtaining an optimal partition of the original network. We then repeat the comparison between FS and $\Delta$FS in terms of their performance after randomly removing an increasing fraction of edges from the original network. In this analysis, we see how $\Delta$FS proves more resilient only for small fractions of edges randomly removed, while for larger values, the two methods both fail in retrieving the original partition of the network. 
Furthermore, the comparison of the two heatmaps of NVI computed between partitions obtained by the same method at different times, in Figures \ref{fig:SBM complex}.c and \ref{fig:teleg_exp_1}.e, suggests how our novel method for the reconstruction of the community structure of a network might uncover relevant scales of the system under study that otherwise would go unnoticed, blurred out due to missing observations.

Throughout our analyses, we only focused on random edge removal with a uniform edge-removal probability, which, in turn, means there is a higher probability of having out-going links removed for nodes with higher out-strength. The positive correlation between the out-strength of nodes and their corresponding uncertainty $\epsilon$, shown in Figure \ref{fig:error-esrimate}.c, suggests that, at least in the case of a messaging platform such as Telegram, this is a realistic scenario.

Lastly, we use in $\Delta$FS the estimates of the number of missing out-going links for each node in the Telegram network, i.e., the uncertainty on the out-strength of nodes, to investigate the differences between the optimal FS partition of the original network and the community structure obtained with $\Delta$FS. 

This work extends the flow-based method for multi-scale community detection in temporal and static directed networks, Flow Stability \cite{doi:10.1126/sciadv.abj3063, Bovet_telegram}, incorporating a method for reconstructing the community structure of a network with missing edges but known experimental error. This last aspect of our novel method prevents us from comparing our analysis to what is obtained with other flow-based community detection methods extended with a similar purpose like those \cite{jelena_unweighted,Jelena_weighted}, where no information regarding the experimental error when \emph{measuring} the network, such as the uncertainty on the out-strength of nodes, is used.
There remain many potential avenues for future work in this area. For instance, in our work, inspired by the function of messaging platforms where users can delete their messages, we modify Flow Stability only to account for the uncertainty on the out-going edges from a node. This leads to the \emph{asymmetric} definition of $\Delta$ Flow Stability in this work, where only the forward diffusive process is modified w.r.t. original Flow Stability. In a scenario where both uncertainties on the out- and in-strength of nodes are known, one could implement a \emph{symmetric version} of $\Delta$ Flow Stability, which we propose in Appendix B. Furthermore, expanding the network reconstruction approach to time-varying networks constitutes a natural further extension of this work. Altogether our findings suggest that $\Delta$ Flow Stability can be a useful tool in empirical settings, where capturing the community structure of directed, weighted networks might be hampered by experimental errors. 

\section{Acknowledgements}
NP and RL received funding from EPSRC Grant Ref EP/V013068/1. 

\section{Data availability}
The original Telegram dataset analysed in this work can be found at \url{https://doi.org/10.7910/DVN/7JUYII}.

\newpage
\section{Appendix A. Approximation with $N\rightarrow\infty$:}
As the computation of the matrix exponential can be relatively time costly for a large network, as in the case of the Telegram network, we introduce a linearization of the transition matrix for the forward process using two linear interpolations:

\begin{equation}
\Tilde{\mathbf{T}}_\text{f}(t) = \begin{cases} (1-t)\mathbf{I} + t \mathbf{M} & \mbox{if } 0 \leq t \leq 1 ,\\ 
\frac{1}{1-t^s}[(t-t^s)\mathbf{M} + (1-t)\mathbf{W}] & \mbox{if } 1 \leq t \leq t^s ,\\ 
\mathbf{W} & \mbox{if } t \geq t^s,
\end{cases}
\end{equation}

where $\mathbf{W}=\lim_{n\rightarrow \infty}\mathbf{M}^n$ is the limiting transition matrix and $t^s$ is the time taken by the random walk to reach stationarity.


We recall the definition of the covariance matrix $\mathbf{S}_\text{f}(t)$:
\begin{equation}
    \mathbf{S}_\text{f}(t)=\mathbf{P}_\text{f}(0)\mathbf{T}_\text{f}(t)\mathbf{T}^\text{inv}_\text{f}(t)-\mathbf{p}_\text{f}(0)^\textsf{T}\mathbf{p}_\text{f}(0)=\mathbf{P}_\text{f}(0)\mathbf{T}_\text{f}(t)\mathbf{P}_\text{f}(t)^{-1}\mathbf{T}^{T}_\text{f}(t)\mathbf{P}_\text{f}(0)-\mathbf{p}_\text{f}(0)^\textsf{T}\mathbf{p}_\text{f}(0).
\end{equation}

We now consider $\mathbf{P}_\text{f}(0)=\mathbf{\Pi}\equiv diag(\pmb{\pi})$ where $\pmb{\pi}$ is the stationary distribution. We then have, when $t\rightarrow0$:

\begin{equation}
    \mathbf{S}_\text{f}(t)\approx\mathbf{\Pi}\Tilde{\mathbf{T}}_\text{f}(t)\mathbf{\Pi}^{-1}\Tilde{\mathbf{T}}^{T}_\text{f}(t)\mathbf{\Pi}-\pmb{\pi}^\textsf{T}\pmb{\pi}
\end{equation}

where the first term on the r.h.s. becomes:

\begin{equation}
    \begin{aligned}
    \mathbf{\Pi}\Tilde{\mathbf{T}}_\text{f}(t)\mathbf{\Pi}^{-1}\Tilde{\mathbf{T}}^{T}_\text{f}(t)\mathbf{\Pi}&=\mathbf{\Pi}\big((1-t)\mathbf{I}+t\mathbf{M}\big)\mathbf{\Pi}^{-1}\big((1-t)\mathbf{I}+t\mathbf{M}^\textsf{T}\big)\big((1-t)\mathbf{I}+t\mathbf{M}\big)\\
    &=\big(\mathbf{\Pi}(1-t)\mathbf{I}\mathbf{\Pi}^{-1}+t\mathbf{\Pi}\mathbf{M}\mathbf{\Pi}^{-1}\big)\big((1-t)\mathbf{I}+t\mathbf{M}^\textsf{T}\big)\mathbf{\Pi}\\
    &=\big((1-t)\mathbf{I}+t\mathbf{\Pi}\mathbf{M}\mathbf{\Pi}^{-1}\big)\big((1-t)\mathbf{I}+t\mathbf{M}^\textsf{T}\big)\mathbf{\Pi}\\
    &= (1-t)^2\mathbf{\Pi}+(1-t)t\big[\mathbf{\Pi}\mathbf{M}+\mathbf{M}^t\mathbf{\Pi}] + t^2\mathbf{\Pi}\mathbf{M}\mathbf{\Pi}^{-1}\mathbf{M}^\textsf{T}\mathbf{\Pi}.
    \end{aligned}
\end{equation}

Neglecting the $t^2$ terms in the expansion, we have:

\begin{equation}
    \mathbf{S}_\text{f}(t) \approx (1-2t)\mathbf{\Pi} + t \big[ \mathbf{\Pi}\mathbf{M}+\mathbf{M}^\textsf{T}\mathbf{\Pi}],
\end{equation}

\NewDocumentCommand{\tens}{t_}
 {%
  \IfBooleanTF{#1}
   {\tensop}
   {\otimes}%
 }
 
where \textbf{M} is defined as:
 \begin{equation}
     \mathbf{M} = \mathbf{I}\tens(1 - \pmb{\alpha})\frac{\mathbf{A}^\text{f}_{\Delta}}{\mathbf{s}_\text{out}}+\frac{t}{\sum_l s^\text{in}_l}\pmb{\alpha}\tens\mathbf{s}_\text{in}.
 \end{equation}

 where $\pmb{\alpha}$ is the 1-D vector encoding the teleportation probabilities of the $N^\text{f}_{err}$ nodes considered in the forward process, and $\mathbf{s}_\text{out}$ and $\mathbf{s}_\text{out}$ are the $N^\text{f}_{err}$ - long vectors of the out- and in-strengths of the nodes, respectively.

Thus, finally:

\begin{equation}
    \mathbf{S}_\text{f}(t)\approx \mathbf{\Pi}\Bigg[(1-2t)\mathbf{I}+t\big[\big(\mathbf{I}\tens(\mathbf{1}-\pmb{\alpha})\frac{\mathbf{A}^\text{f}_{\Delta}}{\mathbf{s}_\text{out}}\big)+\big(\mathbf{I}\tens(\mathbf{1}-\pmb{\alpha})\frac{\mathbf{A}^\text{f}_{\Delta}}{\mathbf{s}_\text{out}}\big)^\textsf{T}\big]\Bigg]+\frac{t}{\sum_l s^\text{in}_l}\mathbf{\Pi}\Bigg[\pmb{\alpha}\tens\mathbf{s}_\text{in}+\mathbf{s}_\text{in}\tens\pmb{\alpha}\Bigg] -\pmb{\pi}^\textsf{T}\pmb{\pi}.
\end{equation}

Analogously, we perform a similar linearisation of the transition matrix of the backward process:

\begin{equation}
\Tilde{\mathbf{T}}_\text{b}(t) = \begin{cases} (1-t)\mathbf{I} + t \mathbf{T}^\text{b}_{\mathbf{DT}} & \mbox{if } 0 \leq t \leq 1 ,\\ 
\frac{1}{1-t^s}[(t-t^s)\mathbf{T}^\text{b}_{\mathbf{DT}} + (1-t)\mathbf{W}_\text{b}] & \mbox{if } 1 \leq t \leq t^s ,\\ 
\mathbf{W}_\text{b} & \mbox{if } t \geq t^s
\end{cases}
\end{equation}

where $\mathbf{T}^\text{b}_{\mathbf{DT}}=\mathbf{I}-\mathbf{L}_\text{b}$ is the discrete-time transition matrix of the backward process; $\mathbf{W}_\text{b}=\lim_{n\rightarrow \infty}(\mathbf{T}^\text{b}_{\mathbf{DT}})^n$ is the limiting transition matrix and $t^s$ is the time taken by the random walk to reach stationarity.

\section{Appendix B. Symmetric Errors}
We propose a generalised version of $\Delta$ Flow Stability that can be employed when both experimental errors on the number of missing outgoing and incoming links per node are known. In this setup, the transition matrices for the forward and backward processes take a similar form to that introduced in Equation (3). We refer here to $\alpha_i$ and $\beta_i$ as the amount of missing out-going and incoming links from/to node $i$, respectively. We then define the transition matrices of the forward as backwards processes, respectively, as:

\begin{equation}
    M^\text{f}_{ij} = (1-\omega_i) \frac{A^\text{f}_{\Delta}(i,j)}{s_\text{out}^i} + \omega_i\frac{F_{ij}}{Z_\beta}\quad,\quad
    M^\text{b}_{ij} = (1-\chi_i) \frac{A^\text{b}_{\Delta}(j,i)}{s_\text{in}^i} + \chi_i\frac{B_{ij}}{Z_\alpha}
\end{equation}
where the teleportation probabilities for the forward process, $\omega$, and for the backward process $\chi$ are defined as follows:
\begin{equation}
\omega_i\equiv\frac{\alpha_i}{\alpha_i+s^i_\text{out}}\quad,\quad
    F_{ij}=\chi_j\quad,\quad
    Z_{\beta}\equiv \sum_{\ell} \chi_\ell
\end{equation}
\begin{equation}
\chi_j\equiv\frac{\beta_j}{\beta_j+s^j_\text{in}}\quad,\quad
    \quad,\quad B_{ij}=\omega_j\quad,\quad
    Z_{\alpha}\equiv \sum_{\ell} \omega_\ell
\end{equation}

    In the forward process, a random walker positioned on node $i$ follows the outgoing links of node $i$ with a certain probability $1-\omega_i$, where $\omega_i$ corresponds to the Dirichlet-rank \cite{Wang2008DirichletRankST} inspired teleportation probability depending on the out-strength of $i$ as well as its \textit{error} $\alpha_i$. If instead the random walker is randomly teleported (with probability $\omega_i$), it can be teleported only to nodes with nonzero error on their in-strength, i.e., to any node $j$ having $\beta_j\neq 0$: the probability of the random walker being teleported from node $i$ to node $j$ is $\chi_j\equiv\frac{\beta_j}{\beta_j+s^j_{in}}$. In the backward process, a random walker positioned on node $i$ follows the ingoing (transpose of the adjacency matrix) links of node $i$ with a certain probability $1-\chi_i$. If instead the random walker is randomly teleported (with probability $\chi_i$), it can be teleported only to nodes with nonzero error on their out-strength, i.e., to any node $j$ having $\alpha_j\neq 0$: the probability of the random walker being teleported from node $i$ to node $j$ is $\omega_j\equiv\frac{\alpha_j}{\alpha_j+s^j_{out}}$.
\bibliographystyle{unsrt}
\bibliography{refs}

\end{document}